
\let\includefigures=\iftrue
\let\useblackboard=\iftrue
\newfam\black
\input harvmac

\noblackbox

\includefigures
\message{If you do not have epsf.tex (to include figures),}
\message{change the option at the top of the tex file.}
\input epsf 
\def\figin{\epsfcheck\figin}\def\figins{\epsfcheck\figins}
\def\epsfcheck{\ifx\epsfbox\UnDeFiNeD
\message{(NO epsf.tex, FIGURES WILL BE IGNORED)}
\gdef\figin##1{\vskip2in}\gdef\figins##1{\hskip.5in}
\else\message{(FIGURES WILL BE INCLUDED)}%
\gdef\figin##1{##1}\gdef\figins##1{##1}\fi}
\def\DefWarn#1{}
\def\figinsert{\goodbreak\midinsert}
\def\ifig#1#2#3{\DefWarn#1\xdef#1{fig.~\the\figno}
B
\writedef{#1\leftbracket fig.\noexpand~\the\figno}%
\figinsert\figin{\centerline{#3}}\medskip\centerline{\vbox{
\baselineskip12pt\advance\hsize by -1truein
\noindent\footnotefont{\bf Fig.~\the\figno:} #2}}
\bigskip\endinsert\global\advance\figno by1}
\else
\def\ifig#1#2#3{\xdef#1{fig.~\the\figno}
\writedef{#1\leftbracket fig.\noexpand~\the\figno}%
\global\advance\figno by1}
\fi
\useblackboard
\message{If you do not have msbm (blackboard bold) fonts,}
\message{change the option at the top of the tex file.}
\font\blackboard=msbm10 scaled \magstep1
\font\blackboards=msbm7
\font\blackboardss=msbm5
\textfont\black=\blackboard
\scriptfont\black=\blackboards
\scriptscriptfont\black=\blackboardss

\else

\fi
%
\def\yboxit#1#2{\vbox{\hrule height #1 \hbox{\vrule width #1
\vbox{#2}\vrule width #1 }\hrule height #1 }}
\def\fillbox#1{\hbox to #1{\vbox to #1{\vfil}\hfil}}
\def\ybox{{\lower 1.3pt \yboxit{0.4pt}{\fillbox{8pt}}\hskip-0.2pt}}
%
%

\def\comments#1{}



\def\II{\relax{I\kern-.10em I}}

\def\IZ{\relax\ifmmode\mathchoice
{\hbox{\cmss Z\kern-.4em Z}}{\hbox{\cmss Z\kern-.4em Z}}
{\lower.9pt\hbox{\cmsss Z\kern-.4em Z}}
{\lower1.2pt\hbox{\cmsss Z\kern-.4em Z}}
\else{\cmss Z\kern-.4emZ}\fi}
\def\IB{\relax{\rm I\kern-.18em B}}
\def\IC{{\relax\hbox{$\inbar\kern-.3em{\rm C}$}}}
\def\ID{\relax{\rm I\kern-.18em D}}
\def\IE{\relax{\rm I\kern-.18em E}}
\def\IF{\relax{\rm I\kern-.18em F}}
\def\IG{\relax\hbox{$\inbar\kern-.3em{\rm G}$}}
\def\IGa{\relax\hbox{${\rm I}\kern-.18em\Gamma$}}
\def\IH{\relax{\rm I\kern-.18em H}}
\def\II{\relax{\rm I\kern-.18em I}}
\def\IK{\relax{\rm I\kern-.18em K}}
\def\IP{\relax{\rm I\kern-.18em P}}

%

\def\inbar{\,\vrule height1.5ex width.4pt depth0pt}

\font\cmss=cmss10 
\def\IR{\relax{\rm I\kern-.18em R}}

%


%

\def\lp10{\ell_p^{10}}
\def\lp11{\ell_p^{11}}
\def\R11{R_{11}}

\def\frac#1#2{{#1 \over #2}}


\hyphenation{Di-men-sion-al}



\lref\toappear{S. Kachru, M. Schulz and E. Silverstein, work in progress.}
\lref\hw{P. Horava and E. Witten, ``Heterotic and Type I String
Dynamics from Eleven-Dimensions,'' Nucl. Phys. {\bf B460} (1996) 506,
hep-th/9510209\semi
E. Witten, ``Strong Coupling Expansion of Calabi-Yau Compactification,''
Nucl. Phys. {\bf B471} (1996) 135, hep-th/9602070\semi
A. Lukas, B. Ovrut, K. Stelle and D. Waldram, ``The Universe as a
Domain Wall,'' Phys. Rev. {\bf D59} (1999) 086001, hep-th/9803235.}

\lref\kaloper{N. Kaloper, ``Bent Domain Walls as Braneworlds,''
Phys. Rev. {\bf D60} (1999) 123506, hep-th/9905210.}

\lref\led{N. Arkani-Hamed, S. Dimopoulos and G. Dvali, ``The Hierarchy
Problem and New Dimensions at a Millimeter,'' Phys. Lett. {\bf B429} (1998)
263, hep-ph/9803315\semi
I. Antoniadis, N. Arkani-Hamed, S. Dimopoulos and G. Dvali, ``New
Dimensions at a Millimeter and Superstrings at a TeV,'' Phys. Lett.
{\bf B436} (1998) 257, hep-ph/9804398.}

\lref\tyezurab{Z. Kakushadze and H. Tye, ``Brane World,'' Nucl. Phys.
{\bf B548} (1999) 180, hep-th/9809147.}

\lref\csaki{C. Csaki, J. Erlich, T. Hollowood and Y. Shirman, ``Universal
Aspects of Gravity Localized on Thick Branes,'' hep-th/0001033.}
\lref\gremm{M. Gremm, ``Four-dimensional Gravity on a Thick Domain
Wall,'' hep-th/9912060.}

\lref\RS{L. Randall and R. Sundrum, ``An Alternative to Compactification,''
Phys. Rev. Lett. {\bf 83} (1999) 4690, hep-th/9906064.} 
\lref\flop{E. Witten, ``Phases of N=2 Theories in Two-Dimensions,''
Nucl. Phys. {\bf B403} (1993) 159, hep-th/9301042\semi
P. Aspinwall, B. Greene and D. Morrison, ``Calabi-Yau Moduli Space,
Mirror Manifolds and Space-time Topology Change in String Theory,''
Nucl. Phys. {\bf B416} (1994) 414, hep-th/9309097.} 
\lref\conifold{A. Strominger, ``Massless Black Holes and Conifolds in
String Theory,'' Nucl. Phys. {\bf B451} (1995) 96, hep-th/9504090.} 
\lref\orbifold{L. Dixon, J. Harvey, C. Vafa and E. Witten, ``Strings on
Orbifolds,'' Nucl. Phys. {\bf B261} (1985) 678.}

\lref\dfgk{O. DeWolfe, D. Freedman, S. Gubser and A. Karch, 
``Modeling the Fifth Dimension with Scalars and Gravity,''
hep-th/9909134.}
\lref\cvetic{
M. Cvetic and H. Soleng, ``Supergravity Domain Walls,''
Phys. Rept. {\bf 282} (1997) 159, hep-th/9604090.}
\lref\nnr{N. Arkani-Hamed, S. Dimopoulos, 
N. Kaloper and R. Sundrum, to appear.}
\lref\rutgers{O. Aharony, T. Banks, A. Rajaraman and M. Rozali, unpublished 
ideas.}
\lref\rubshap{V. Rubakov and M. Shaposhnikov, ``Extra Space-Time Dimensions: 
Towards a Solution to the Cosmological Constant Problem," Phys. Lett.
{\bf B125} (1983) 139.} 
\lref\verlinde{E. Verlinde and H. Verlinde, ``On RG Flow and the
Cosmological Constant,'' hep-th/9912058.}
\lref\schmidhuber{C. Schmidhuber, ``AdS(5) and the 4d Cosmological
Constant,'' hep-th/9912156.}
\lref\adscft{J. Maldacena, ``The Large N Limit of Superconformal
Field Theories and Supergravity,'' Adv. Theor. Math. Phys. {\bf 2} (1998)
231, hep-th/9711200\semi
S. Gubser, I. Klebanov and A. Polyakov, ``Gauge Theory Correlators
from Noncritical String Theory,'' Phys. Lett. {\bf B428} (1998) 105,
hep-th/9802109\semi
E. Witten, ``Anti-de Sitter Space and Holography,'' Adv. Theor. Math.
Phys. {\bf 2} (1998) 253, hep-th/9802150.}
\lref\kss{S. Kachru, N. Seiberg and E. Silverstein, ``SUSY Gauge
Dynamics and Singularities of 4d N=1 String Vacua,'' Nucl. Phys. 
{\bf B480} (1996) 170, hep-th/9605036\semi
S. Kachru and E. Silverstein, ``Singularities, Gauge Dynamics and
Nonperturbative Superpotentials in String Theory,'' Nucl. Phys.
{\bf B482} (1996) 92, hep-th/9608194.}
\lref\ksorb{S. Kachru and E. Silverstein, ``4d Conformal Field Theories
and Strings on Orbifolds,'' Phys. Rev. Lett. {\bf 80} (1998) 4855, 
hep-th/9802183.}

\lref\oldjoe{S. de Alwis,  J. Polchinski and R. Schimmrigk, ``Heterotic
Strings with Tree Level Cosmological Constant,'' Phys. Lett. 
{\bf B218} (1989) 449.}
\lref\kks{S. Kachru, J. Kumar and E. Silverstein, ``Orientifolds,
RG Flows, and Closed String Tachyons,'' hep-th/9907038.}
\lref\smallinst{E. Witten, ``Small Instantons in String Theory,''
Nucl. Phys. {\bf B460} (1996) 541, hep-th/9511030.}

\lref\nati{
N. Seiberg,
``Matrix Description of M-theory on $T^5$ and $T^5/Z_2$,'' 
Phys.Lett. {\bf B408} (1997) 98, hep-th/9705221.}
\lref\ntwo{A. Klemm, W. Lerche, P. Mayr, C.Vafa and N. Warner,
``Self-Dual Strings and N=2 Supersymmetric Field Theory,''
Nucl.Phys. {\bf B477} (1996) 746, hep-th/9604034.}
\lref\perletc{N. Bahcall, J. Ostriker, S. Perlmutter and P. Steinhardt,
``The Cosmic Triangle: Assessing the State of the Universe,'' Science
{\bf 284} (1999) 1481, astro-ph/9906463.
}
\lref\polwitt{J. Polchinski and E. Witten, 
``Evidence for Heterotic - Type I String Duality,''
Nucl.\ Phys.\  {\bf B460} (1996) 525,
hep-th/9510169.}

\Title{\vbox{\baselineskip12pt\hbox{hep-th/0001206}
\hbox{SLAC-PUB-8337}\hbox{SU-ITP-00/02}\hbox{IASSNS-HEP-00/05}}}
{\vbox{
\centerline{Self-tuning flat domain walls} 
\medskip
\centerline{in 5d gravity and string theory} }}
\bigskip 
\centerline{Shamit Kachru, Michael Schulz and Eva Silverstein}
\bigskip
\centerline{Department of Physics and SLAC}
\smallskip
\centerline{Stanford University}
\smallskip
\centerline{Stanford, CA 94305/94309}

\bigskip
\noindent

We present Poincare invariant domain wall (``3-brane'') 
solutions to some 5-dimensional
effective theories which can arise naturally in string theory.
In particular, we find theories where Poincare invariant
solutions exist for arbitrary values of the brane tension, for
certain restricted forms of the bulk interactions. 
We describe examples in string theory where it would be natural
for the quantum corrections to the tension of the brane
(arising from quantum fluctuations
of modes with support on the brane)
to maintain the required form of the action.  In such cases,
the Poincare invariant solutions persist in the
presence of these
quantum corrections
to the brane tension, so that no 4d cosmological constant is
generated by these modes.

\Date{January 2000}

\newsec{Introduction}

Some time ago, it was suggested that the cosmological constant problem
may become soluble in models where our world is a topological defect 
in some higher dimensional spacetime \rubshap.
Recently such models have come under renewed investigation. 
This has been motivated
both by brane world scenarios (see for instance
\refs{\hw,\led,\tyezurab}) and by the suggestion of Randall
and Sundrum \RS\ that the four-dimensional graviton might be a bound
state of a 5d graviton to a 4d domain wall.
At the same time, new ideas relating 4d renormalization group flows 
to 5d AdS gravity via the AdS/CFT correspondence \adscft\ have 
inspired related approaches to explaining 
the near-vanishing of
the 4d cosmological term \refs{\verlinde,\schmidhuber}.  
These authors suggested (following \rubshap) that quantum
corrections to the 4d cosmological constant could
be cancelled by variations of fields in a five-dimensional
bulk gravity solution.  The results of this paper might be 
regarded as a concrete partial 
realization of this scenario, in the context of 
5d dilaton gravity and string theory.
A different AdS/CFT motivated approach to this problem appeared in 
\ksorb.

In the thin wall approximation, we can represent a domain wall
in 5d gravity by a delta function source with some coefficient 
$f(\phi)$ (where $\phi$ is a bulk scalar field, the dilaton),
parametrizing the tension of the wall.  Quantum 
fluctuations of the fields with support on the brane
should correct 
$f(\phi)$. 
In this paper, we present  
a concrete example of a 5d dilaton gravity theory where
one can find Poincare invariant domain wall solutions 
for $\it generic$ $f(\phi)$.  The constraint of finding
a finite 4d Planck scale then restricts 
the sign of $f$ and the value
of ${f^\prime\over f}$ at the wall to lie
in a range of order one.  Thus fine-tuning is
not required in order to avoid having the quantum fluctuations
which correct $f(\phi)$ generate a 4d cosmological
constant.  
One of the requirements we must impose is
that the 5d cosmological constant $\Lambda$ should 
vanish.\foot{It is possible that an Einstein frame bulk cosmological
term which is independent of $\phi$ will also allow for similar physics
\toappear.} 
This would
be natural in scenarios where the bulk is supersymmetric (though the
brane need not be), or where quantum corrections to the bulk are
small enough to neglect in a controlled expansion.

For suitable choices of
$f(\phi)$, this example exhibits
the precise dilaton couplings which naturally
arise in string theory.  There are two interesting and
distinct contexts in which this happens.    
One is to consider $f(\phi)$ corresponding to
tree-level dilaton coupling ($V e^{-2\phi}$ in string
frame, for some constant $V$).  This 
form of the dilaton coupling is
not restricted to tree-level $\it perturbative$ string
theory -- it occurs
for example on the worldvolumes of $NS$ branes in
string theory.  There, the dynamics of the worldvolume
degrees of freedom does not depend on the dilaton -- the
relevant coupling constant is dilaton independent.  
Therefore, quantum corrections to the brane tension
due to dynamics of worldvolume fields
would be expected to maintain the ``tree-level'' form
of $f(\phi)$, while simply shifting the coefficient
$V$ of the (string frame) $e^{-2\phi}$.
The other form of
$f(\phi)$ natural in string theory involves a
power series in $e^\phi$.  This type of coupling
occurs when quantum corrections are controlled
by the dilaton in string theory.

In either case,
as long as we only consider quantum corrections which modify 
$f(\phi)$ but maintain
the required form of the bulk 5d gravity action, this 
means that quantum corrections to the brane tension do not
destabilize flat space; they do not 
generate a
four-dimensional cosmological constant.  
We will argue that
some of our examples should have a microscopic realization
in string theory with this feature, at leading
order in a controllable approximation scheme. 
It is perhaps appropriate to call this ``self-tuning'' of
the cosmological constant because the 5d gravity theory and its
matter fields respond in just the right way to shifts in the
tension of the brane to maintain 4d Poincare invariance.
Note that here, as in \RS, there is a distinction between
the brane tension and the 4d cosmological constant.

There are two aspects of the  
solutions we find which are not under
satisfactory control.
Firstly, the curvature in the brane solutions of interest has singularities
at finite distance from the wall; the proper interpretation of these
singularities will likely be crucial to understanding the mechanism
of self-tuning from a four-dimensional perspective.
We cut off the space at
these singularities.  The wavefunctions for the four-dimensional
gravitons in our solutions vanish there.
Secondly, the value of the dilaton $\phi$ diverges at some of 
the singularities;
this implies that the theory is becoming strongly coupled there.
However, the curvature and coupling can be kept arbitrarily weak 
at the core of the wall.  Therefore, some aspects of the solutions
are under control and we think the self-tuning mechanism can be
concretely studied.  We present some preliminary ideas
about the microscopic nature of the singularities in \S3.   

A problem common to the system studied here and that
of \RS\ is the possibility of instabilities, hidden
in the thin wall sources, that are missed by the
effective field theory analysis.  
Studying thick wall analogues of our solutions would
probably shed light on this issue. 
We do not resolve
this question here.  But taking advantage of the 
stringy dilaton couplings possible in our set
of self-tuned models, we present a plausibility
argument for the existence of stringy realizations,
a subject whose details we leave for future work \toappear.

Another issue involves solutions where the wall is
not Poincare invariant.  This could mean it is curved (for
example, de Sitter or
Anti de Sitter).  However
it could also mean 
that there is a nontrivial dilaton profile along the wall
(one example being the linear dilaton solution in string theory, which 
arises 
when the tree-level cosmological constant
is nonvanishing).  This latter possibility is a priori as likely as others,
given the presence of the massless dilaton in our solutions. 

Our purpose
in this paper is to argue that starting with a Poincare invariant wall,
one can find systems where quantum corrections 
leave a Poincare invariant wall as a solution.  However 
one could also imagine
starting with non Poincare invariant wall solutions of
the same 5d equations (and preliminary analysis suggests 
that such solutions do exist in the generic case, 
with finite 4d Planck scale). 
We are in the process of systematically 
analyzing the fine tuning of initial conditions
that considering a classically Poincare invariant wall might entail
\toappear.

The paper is organized as follows.
In \S2, we write down the 5d gravity + dilaton theories that we
will be investigating.  
We solve the equations of motion
to find Poincare invariant domain walls,
both in the cases where the 5d Lagrangian has couplings which
provide the self-tuning discussed above, and in more general cases.
In \S3, we describe several possible embeddings of our results
into a more microscopic string theory context.
We close with a discussion of promising directions for future
thought in \S4.

There have been many interesting recent papers which study domain walls in
5d dilaton gravity theories.  We particularly found \dfgk\ and
\cvetic\ useful, and further references may be found there. 

This research was inspired by very interesting discussions with
O. Aharony and T. Banks. 
While our work on Poincare invariant domain walls and self-tuning 
was in progress, we learned that very similar work
was in progress by Arkani-Hamed, Dimopoulos,
Kaloper and Sundrum \nnr.  
In particular, before we had obtained the solutions
in \S2.3 and \S2.4, R. Sundrum told us that 
they were finding singular solutions to
the equations and were hoping the singularities would ``explain''
a breakdown of 4d effective field theory on the domain wall.

\newsec{Poincare-invariant 4d Domain Wall Solutions}

\subsec{Basic Setup and Summary of Results}

Let us consider the action

\eqn\basicac{\eqalign{
S=
&\int d^5x\sqrt{-G}\biggl[
R-{4\over{3}}(\nabla\phi)^2-\Lambda e^{a\phi}\biggr]\cr
&+\int d^4x\sqrt{-g}(-f(\phi))\cr
}
}
describing a scalar field $\phi$ and gravity living
in five dimensions coupled to a thin four-dimensional
domain wall.  Let us set the position of the domain
wall at $x_5=0$.  
Here we follow the notation of \RS\ so that the metric
$g_{\mu\nu}$ along the four-dimensional slice at $x_5=0$ is given 
in terms of the five-dimensional metric $G_{MN}$ by 
\eqn\metfour{\eqalign{
&g_{\mu\nu}=\delta_\mu^M\delta_\nu^NG_{MN}(x_5=0)\cr
&\mu,\nu=1,\dots,4\cr
&M,N=1,\dots,5\cr
}
}

For concreteness, in much of our discussion we will make the
choice
\eqn\dilcho{f(\phi) = V e^{b\phi}}
However, most of our considerations will $\it not$ depend on this
detailed choice of $f(\phi)$ (for reasons that will become clear).
With this choice, \basicac\
describes a family of theories parameterized by 
$V$, $\Lambda$, $a$, and $b$.  If $a=2b=4/3$, the
action \basicac\ agrees with tree-level string
theory where $\phi$ is identified with the dilaton.
(That is, the 5d cosmological constant term
and the 4d domain wall tension term both scale
like $e^{-2\phi}$ in string frame.)  
In \S3\ we will
discuss a very natural context in which this type
of action arises in string theory, either with the specific
form \dilcho\ or with more general $f(\phi)$.

In the rest of this section we will derive
the field equations arising from this action
and construct some interesting solutions of these
equations.
In particular, we will be interested in whether
there are Poincare-invariant solutions
for the metric of the four-dimensional slice at $x_5=0$
for generic values of these parameters (or more
generally, for what subspaces of this parameter space
there are Poincare-invariant solutions in four dimensions).    
We will also require that the geometry is such
that the four-dimensional Planck scale is
finite.  
Our main results can be summarized in three different cases as follows:
\bigskip

\noindent (I)  For $\Lambda=0$, $b\ne\pm{4\over 3}$ but
otherwise arbitrary, and arbitrary magnitude of
$V$ we find a Poincare-invariant domain wall solution of the
equations of motion.  For $b=2/3$, which is the value corresponding
to a brane tension of order $e^{-2\phi}$ in string frame,
the sign of $V$ must be positive in order to
correspond to a solution with a finite four-dimensional
Planck scale, but it is otherwise unconstrained.
This suggests that for fixed scalar field coupling to the
domain wall, quantum corrections to its tension $V$ do
not spoil Poincare invariance of the slice.
In \S3\ we will review examples
in string theory of situations where worldvolume
degrees of freedom contribute
quantum corrections to the $e^{-2\phi}$ term in
a brane's tension.  Our result implies that these quantum
corrections do not need to be fine-tuned to zero to
obtain a flat four-dimensional spacetime.  

For a generic choice of $f(\phi)$ in \basicac\ (including
the type of power series expansion in $e^{\phi}$ that would
arise in perturbative string theory), the same basic
results hold true:  We are able to find Poincare invariant
solutions without fine-tuning $f$.  Insisting on a 
finite 4d Planck scale gives a furthur constraint on
$f^\prime/f$ at the wall, forcing it to lie
in a range of order one.     

Given a solution with one value of $V$ and $\Lambda=0$,
a self-tuning mechanism is in fact clear from
the Lagrangian (for $b \neq 0$).  In \basicac\ we see that
if $\Lambda=0$ (or $a=0$), the only non-derivative
coupling of the dilaton is to the brane tension
term, where it appears in the combination $(-V)e^{b\phi}$.  
Clearly given a solution for one value of $V$, there
will be a solution for any value of $V$ obtained by absorbing
shifts in $V$ with shifts in $\phi$.  With more general
$f(\phi)$, similar remarks hold: the dilaton zero mode
appears only in $f$, and 
one can absorb
shifts in $V$ by shifting $\phi$.     

However, in the special case $b=0$ (where $f(\phi)$ is just a constant),
we will also find flat solutions for generic $V$.  This implies
that the freedom to vary 
the dilaton zero mode is not the only mechanism that ensures
the existence of a flat solution for arbitrary $V$.

\bigskip
\noindent (II)  For $\Lambda=0$, $b=\pm {4/3}$, we find
a different Poincare-invariant solution
(obtained by matching together two 5d bulk
solutions in a different combination than
that used in obtaining the solutions described
in the preceding paragraph (I)).  A solution is present
for any value of $V$.
This suggests that for fixed
scalar field coupling to the domain wall, quantum 
corrections to its tension $V$ do not spoil Poincare-invariance
of the slice.  Again the sign of $V$ must be positive
in order to have a finite four-dimensional Planck scale.  

\bigskip
\noindent (III)  We do not find a solution (nor
do we show that none exists) for general
$\Lambda$, $V$, $a$, and $b$ (in concordance with
the counting of parameters in \dfgk).  However, for
each $\Lambda$ and $V$ there is a choice of
$a$ and $b$ for which we do find a Poincare invariant
solution using a simple ansatz.  

For $a=0$, and general $b$,
$\Lambda$, and $V$ we are currently investigating the existence
of self-tuning solutions.  Their existence would be 
in accord with the fact that in this case,
as in the cases with $\Lambda=0$, the dilaton zero
mode only appears in the tension of the wall.
This means again that
shifts in $V$ can be absorbed
by shifting $\phi$, so if one finds a Poincare invariant solution
for any $V$, one does not need to fine-tune $V$ to solve the equations.

\subsec{Equations of Motion}

The equations of motion arising for the theory \basicac, with
our simple choice for $f(\phi)$ given in \dilcho, are
as follows.  Varying with respect to the dilaton gives:
\eqn\dileq{
\sqrt{-G}\biggl({8\over 3}\nabla^2\phi-a\Lambda e^{a\phi}\biggr)
-bV\delta(x_5)e^{b\phi}\sqrt{-g}
=0
}
The Einstein equations for this theory are:
\eqn\eineqs{\eqalign{
&\sqrt{-G}\biggl(R_{MN}-{1\over 2}G_{MN}R\biggr)\cr
&-{4\over 3}\sqrt{-G}\biggl[\nabla_M\phi\nabla_N\phi
-{1\over 2}G_{MN}(\nabla\phi)^2\biggr]\cr
&+{1\over 2}\biggl[\Lambda e^{a\phi}\sqrt{-G}G_{MN}
+\sqrt{-g}Vg_{\mu\nu}\delta^\mu_M\delta^\nu_N\delta(x_5)
\biggr]=0\cr
}
}

We are interested in whether there are
solutions with Poincare-invariant four-dimensional
physics.  Therefore we look for solutions of \dileq\ and
\eineqs\ where the metric takes the form
\eqn\metans{
ds^2=e^{2A(x_5)}(-dx_1^2+dx_2^2+dx_3^2+dx_4^2)+dx_5^2
}

With this ansatz for the metric, the equations become
\eqn\firsteq{
{8\over 3}\phi''+{32\over 3}A'\phi'-a\Lambda e^{a\phi}
-bV\delta(x_5)e^{b\phi}
= 0
}
\eqn\secondeq{
6(A')^2-{2\over 3}(\phi')^2+{1\over 2}\Lambda e^{a\phi}
= 0
}
\eqn\thirdeq{
3A''+{4\over 3}(\phi')^2+{1\over 2}e^{b\phi}V\delta(x_5)
= 0}
where $'$ denotes differentiation with respect to
$x_5$.  
The first one \firsteq\ is the dilaton equation of motion,
the second \secondeq\ is the 55 component of Einstein's
equations, and the last \thirdeq\ comes from a linear combination
(the difference) of the $\mu\nu$ 
component of Einstein's equation and the 55 component.  

We will mostly consider the simple ansatz
\eqn\ansatz{
A'=\alpha\phi'.
}
However for the case $a=0$, $\Lambda\ne 0$ we will integrate
the equations directly.

\subsec{$\Lambda=0$ Case}

Let us first consider the system with $\Lambda=0$.
We will first study the bulk equations of motion
(i.e. the equations of motion away from $x_5=0$) where
the $\delta$-function terms in \firsteq\ and \thirdeq\ do
not come in.  
Note that because the delta function terms do not enter,
the bulk equations are independent of our choice of
$f(\phi)$ in \basicac.
We will then consider the conditions
required to match two bulk solutions on either
side of the domain wall of tension $Ve^{b\phi}$ at
$x_5=0$.  We will find two qualitatively different
ways to do this, corresponding to results (I) and (II)
quoted above.  We will also find that for fairly generic $f(\phi)$,
the same conclusions hold.    

\bigskip
\centerline{\it Bulk Equations:  $\Lambda=0$}
 
\bigskip
 
Plugging the ansatz \ansatz\ into \secondeq\
(with $\Lambda=0$) we find that
\eqn\secondsimpl{6\alpha^2(\phi')^2={2\over 3}(\phi')^2}
which is solved if we take 
\eqn\solalph{
\alpha=\pm{1\over 3}
}
Plugging this ansatz into \firsteq\ we obtain
\eqn\firstagain{
{8\over 3}(\phi''+4(\pm{1\over 3})(\phi')^2)=0
}
Plugging it into \thirdeq\ we obtain
\eqn\thirdagain{
3(\pm{1\over 3})\phi''+{4\over 3}(\phi')^2=0
}
With either choice of sign for $\alpha$, these
two equations become identical in bulk.  For
$\alpha=\pm{1\over 3}$, we must solve
\eqn\pluseq{
\phi''\pm{4\over 3}(\phi')^2=0
}
in bulk.  
This is solved by
\eqn\solphi{
\phi=\pm{3\over 4}\log|{4\over 3}x_5+c|+d
}
where $c$ and $d$ are arbitrary integration constants.

Note that there is a singularity in this solution
at 
\eqn\singpoint{x_5=-{3\over 4}c
}
Our solutions will involve
regions of spacetime to one side of this singularity;
we will assume that it
can be taken to effectively cut off the space.  
At present we do not have much quantitative to
say about the physical implications of this
singularity.   The results we derive
here (summarized above) strongly motivate
further exploring the effects of these singularities
on the four-dimensional physics of our domain
wall solutions.  

At $x_5=0$ there is localized energy density leading
to the $\delta$-function terms in \firsteq\ and \thirdeq.  
We can solve these equations by introducing appropriate
discontinuities in $\phi'$ at the wall (while insisting
that $\phi$ itself is continuous).  We will now
do this for two illustrative cases (the first being
the most physically interesting).

\bigskip
\centerline{\it Solution (I):}  
\bigskip

Let us take the bulk solution with $\alpha=+{1\over 3}$
for $x_5<0$, and the one with $\alpha=-{1\over 3}$ for
$x_5>0$.  So we have
\eqn\firstreg{
\phi(x_5) = \phi_1(x_5)={3\over 4}\log|{4\over 3}x_5+c_1|+d_1, ~~~~x_5<0
}
\eqn\secondreg{
\phi(x_5) = \phi_2(x_5)=-{3\over 4}\log|{4\over 3}x_5+c_2|+d_2, ~~~~x_5>0
}
where we have allowed for the possibility that the (so far) arbitrary
integration constants can be different on the two
sides of the domain wall.      

Imposing continuity of $\phi$ at $x_5=0$ leads to 
the condition
\eqn\matchphi{
{3\over 4}\log|c_1|+d_1=-{3\over 4}\log|c_2|+d_2
}
This equation determines the
integration constant $d_2$ in terms of the others.

To solve \firsteq\ we then require
\eqn\firstmatch{
{8\over 3}(\phi_2^\prime(0)-\phi_1^\prime(0))=bVe^{b\phi(0)}
}
while to solve \thirdeq\ we need
\eqn\thirdmatch{
3\biggl(\alpha_2\phi_2^\prime(0)-\alpha_1\phi_1^\prime(0)\biggr)
=-{1\over 2}Ve^{b\phi(0)}
}
(where $\alpha_1=+{1\over 3}$ and $\alpha_2=-{1\over 3}$).
These two matching conditions become
\eqn\firstmtwo{
-{8\over 3}({1\over c_1}+{1\over c_2})
=bVe^{bd_1}|c_1|^{{3\over 4}b}
}
and
\eqn\thirdmtwo{
{1\over c_2}-{1\over c_1}=
-{1\over 2}Ve^{bd_1}|c_1|^{{3\over 4}b}
}
Solving for the integration constants
$c_1$ and $c_2$ we find
\eqn\ctwo{
{2\over c_2}=\biggl[-{{3b}\over 8}-{1\over 2}\biggr]
Ve^{bd_1}|c_1|^{{3\over 4}b}
}
\eqn\cone{
{2\over c_1}=\biggl[-{{3b}\over 8}+{1\over 2}\biggr]
Ve^{bd_1}|c_1|^{{3\over 4}b}
}

Note that as long as $b\ne\pm{4\over 3}$, we here find
a solution for the integration constants $c_1$ and $c_2$
in terms of the parameters $b$ and $V$ which appear
in the Lagrangian and the integration constant $d_1$.  
(As discussed above, the integration constant $d_2$ is
then also determined).\foot{
We will momentarily find a disjoint set of $\Lambda = 0$ domain
wall solutions for which $b$ will be forced to
be $\pm 4/3$, so altogether there are solutions for any $b$.}  In 
particular, for
scalar coupling given by $b$, there is a Poincare-invariant
four-dimensional domain wall for any value of the
brane tension $V$; $V$ does not need to be fine-tuned
to find a solution.  As is clear from the form of
the 4d interaction in \basicac, one way to understand this 
is that the scalar field $\phi$ can absorb
a shift in $V$ since the only place that 
the $\phi$ zero mode appears in the Lagrangian is 
multiplying $V$.   However since we can use these
equations to solve for $c_{1,2}$ without fixing $d_1$,
a more general story is at work; in particular, even
for $b=0$ we find solutions for arbitrary $V$. 

A constraint on the sign of $V$ arises, as we
will now discuss, from the requirement that
there be singularities \singpoint\ in the bulk
solutions, effectively cutting off the 
$x_5$ direction at finite volume.  

\medskip
\noindent {\it More General $f(\phi)$} 
\medskip

If instead of \dilcho\ we include a more general choice of
$f$ in the action \basicac, the considerations above go through
unaltered.  The choice of $f$ only enters in the matching
conditions \firstmatch\ and \thirdmatch\ at the domain wall.
The modified equations become

\eqn\firstmod{{8\over 3} (\phi^{\prime}_{2}(0) - \phi^{\prime}_{1}(0))
= {\partial f \over \partial \phi}(\phi(0))}
\eqn\thirdmod{3\biggl( 
\alpha_2 \phi^{\prime}_{2}(0) - \alpha_{1} \phi^{\prime}_{1}(0)
\biggr) = -{1\over 2}f(\phi(0))} 

\noindent
In terms of the integration constants, these become:

\eqn\firstint{-{8\over 3}({1\over c_1} + {1\over c_2}) = {\partial f \over
\partial \phi}({3\over 4} \log\vert c_1\vert + d_1)}
\eqn\secint{{1\over c_2} - {1\over c_1} = -{1\over 2} f 
({3\over 4}\log \vert c_1\vert + d_1)} 
Clearly for generic $f(\phi)$, one can solve these equations.

\medskip
\noindent {\it Obtaining a Finite 4d Planck Scale}
\medskip

Consider the solution \firstreg\ on
the $x_5<0$ side.  If 
$c_1<0$, then there is never a singularity.
Let us consider the four-dimensional Planck
scale.  It is proportional to the integral \RS\
\eqn\fourPlanck{
\int dx_5 ~e^{2A(x_5)}
}
In the $x_5<0$ region, this goes like
\eqn\fourus{
\int dx_5 \sqrt{|{4\over 3}x_5+c_1|}
}

If $c_1<0$, then there is no singularity, and
this integral is evaluated from $x_5=-\infty$
to $x_5=0$.  It diverges.
If $c_1>0$, then there is a singularity at
\singpoint.  Cutting off the volume integral
\fourus\ there gives a finite result.  Note that the
ansatz \ansatz\ leaves an undetermined integration
constant in  
$A$, so one can tune the actual value of the 4d
Planck scale by shifting this constant. 

In order to have a finite 4d Planck scale, we
therefore impose that $c_1>0$.  This requires
$V({1\over 2}-{{3b}\over 8})>0$.  For the
value $b=2/3$, natural in string theory
(as we will discuss in \S3), this requires
$V>0$.  With this constraint, there 
is similarly a singularity on the
$x_5>0$ side which cuts off the volume on
that side.    

These conditions extend easily to conditions on
$f(\phi)$ in the more general case.  We find
\eqn\finitef{\eqalign{
& -{3\over 8}{{\partial f}\over{\partial\phi}}(\phi(0))-
   {1\over 2}f(\phi(0))<0\cr
&  -{3\over 8}{{\partial f}\over{\partial\phi}}(\phi(0))+
   {1\over 2}f(\phi(0))>0\cr
  }
}
This means that $f(\phi)$ must be positive at the wall
(corresponding to a positive tension brane),
and that
\eqn\range{
-{4\over 3}<{f^\prime\over f}<{4\over 3}
}
So although $f$ does not need to be fine-tuned to
achieve a solution of the sort we require, it
needs to be such that $f^\prime/f$ is in the range
\range.  

Let us discuss some of the physics at the singularity.
Following \refs{\RS,\dfgk}, we can compute the $x_5$-dependence
of the four-dimensional graviton wavefunction.  
Expanding the metric about our solution
(taking $g_{\mu\nu}=e^{2A}\eta_{\mu\nu}+h_{\mu\nu}$),
we find 
\eqn\waveftn{
h_{\mu\nu}\propto \sqrt{|{4\over 3}x_5+c|}
}
At a singularity, where $\vert {4\over 3}x_5+c\vert$ vanishes,
this wavefunction also vanishes.  Without understanding
the physics of the singularity, we cannot determine
yet whether it significantly affects the interactions
of the four-dimensional modes.  

It is also of interest to consider the behavior of
the scalar $\phi$ at the singularities.  In string
theory this determines the string coupling.
In our solution (I), we see that 
\eqn\dilsing{\eqalign{
& x_5\to -{3\over 4}c_1  \Rightarrow  \phi\to -\infty \cr
& x_5\to -{3\over 4}c_2   \Rightarrow  \phi\to  \infty \cr
}}
So in string theory, the curvature singularity on the
$x_5<0$ side is weakly coupled, while that on
the $x_5>0$ side is strongly coupled.  It may be
possible to realize these geometries in a context
where supersymmetry is broken by the brane, so
that the bulk is supersymmetric.  In such a case
the stability of the high curvature 
and/or strong-coupling regions may be easier to 
ensure.  In any case we believe that the results
of this section motivate further analysis of
these singular regions, which we leave for future work.

Putting everything together, we have found
the solution described in case (I) above.
It should be 
clear that since $f(\phi)$ only appears in \basicac\
multiplying the delta function ``thin wall'' source term, 
we can always use the choice \dilcho\ in writing matching
conditions at the wall for concreteness.  To understand what
would happen with a more general $f$, one simply replaces
$Ve^{b\phi(0)}$ with $f(\phi(0))$ 
and $bVe^{b\phi(0)}$ with ${\partial f \over \partial \phi}(\phi(0))$ 
in the matching equations.
We will not explicitly say this in each case, but it makes the
generalization to arbitrary $f$ immediate.

\bigskip
\centerline{\it Solution (II):}
\bigskip

A second type of solution with $\Lambda=0$ is
obtained by taking $\alpha$ to have the same
sign on both sides of the domain wall.  
So we have 
\eqn\firstregII{
\phi(x_5) = \phi_1(x_5)=\pm{3\over 4}\log|{4\over 3}x_5+c_1|+d_1, ~~~~x_5<0
}
\eqn\secondregII{
\phi(x_5) = \phi_2(x_5)=\pm{3\over 4}\log|{4\over 3}x_5+c_2|+d_2, ~~~~x_5>0  
}
The matching conditions then require $b=\mp {4\over 3}$
for consistency of \firsteq\ and \thirdeq\ (in the case with more
generic $f(\phi)$, this generalizes to the condition 
${\partial f\over \partial \phi}(\phi(0)) = \mp {4\over 3} f(\phi(0))$).  
This
is not a value of $b$ that appears from a dilaton
coupling in perturbative string theory.  It is
still interesting, however, as a gravitational low-energy
effective field theory where $V$ does not have to
be fine-tuned in order to preserve four-dimensional
Poincare invariance.  We find a solution to
the matching conditions with
\eqn\dsol{\eqalign{
& c_1=c,~~~~x_5>0 \cr
& c_2=-c,~~~~x_5<0\cr
& d_1=d_2=d\cr
&e^{\mp{4\over 3}d}= {4\over V} {c\over \vert c\vert}\cr
}
}
for some arbitrary constant $c$, and any $V$.
This gives the results summarized in case (II) above.
The value $b=\mp 4/3$, which is required here, 
was excluded from the solutions
(I) derived in the last section. 

As long as we choose $c$ such that there are singularities on
both sides of the domain wall, we again get finite 4d Planck
scale.   
As we can see from \firstregII\ and \secondregII, having singularities
on either side of the origin requires $c$ to be positive. 
Then we see from \dsol\ that we can find a solution for arbitrary
positive brane tension $V$.

Let us discuss the physics of the singularities in this case.
As in solutions (I), the graviton wavefunction decays to
zero at the singularity like $(x-x_{sing})^{1\over 2}$.  
For $b=-{4/3}$, $\phi\to -\infty$ at the singularities
on both sides, while for $b={4\over 3}$, $\phi\to\infty$
at the singularities on both sides.  

Putting solutions (I) and (II) together, we see that in the
$\Lambda = 0$ case one can find a Poincare invariant solution
with finite 4d Planck scale for any positive tension $V$ and
any choice of $b$ in \basicac.  
As we have seen, this in fact remains true with \dilcho\ replaced
by a more general dilaton dependent brane tension $f(\phi)$.

\bigskip
\centerline{\it  Two-Brane Solutions}
\bigskip

One can also obtain solutions describing a pair
of domain walls localized in a compact fifth dimension.
In case (I), one can show that such solutions 
always involve singularities.  In case (II), there are 
solutions which avoid singularities while maintaining
the finiteness of the four-dimensional Planck
scale.  They however involve extra
moduli (the size of the compactified fifth
dimension) which may be stabilized
by for example the mechanism of
\lref\gw{W. Goldberger and M. Wise, ``Modulus Stabilization with
Bulk Fields,'' Phys. Rev. Lett. {\bf 83} (1999) 4922, hep-ph/9907447.}
\gw.  The singularity is avoided in these cases by placing
a second domain wall between $x_5=0$ and
the would-be singularity at ${4\over 3}x_5+c=0$.
This allows us in particular to find solutions
for which $\phi$ is bounded everywhere, so that
the coupling does not get too strong.       
This is a straightforward generalization of
what we have already done and we will not 
elaborate on it here.  

\subsec{$\Lambda\ne 0$ (Solution III)}

More generally we can consider the entire Lagrangian
\basicac\ with parameters $\Lambda, V, a, b$.  
In this case, plugging in the ansatz \ansatz\ to
equations \firsteq --\thirdeq, we find a bulk solution
\eqn\gensol{\eqalign{
&\phi=-{2\over a}\log({{a(\mp\sqrt{B})}\over 2}x_5+d)\cr
&B={\Lambda\over{{{4\over 3}-12\alpha^2}}}\cr
&\alpha=-{8\over{9a}}\cr
}
}
We find a domain wall solution by taking one sign
in the argument of the logarithm 
in \gensol\ for
$x_5<0$ and the opposite sign in the argument of
the logarithm for $x_5>0$.  Say for instance that $a>0$.  
Then we could take the $-$ sign for $x>0$ and the $+$
sign for $x<0$, and find a solution which terminates
at singularities on both sides if we choose $d>0$.
  
The matching conditions then require
\eqn\lastmatch{
V=-12\alpha\sqrt{B}
}
and
\eqn\bans{
b=-{4\over{9\alpha}}
}

So we see that here $V$ must be fine-tuned to
the $\Lambda$-dependent value given in \lastmatch.
This is similar to the situation in \RS, where
one fine-tune is required to set the four-dimensional
cosmological constant to zero.  Like in
our solutions in \S2.1, there is one undetermined 
parameter in the Lagrangian.  But here it
is a complicated combination of $\Lambda$ and
$V$ (namely, ${V\over \sqrt{\Lambda}}$), and 
we do not have an immediate interpretation
of variations of this parameter as arising from
nontrivial quantum corrections from a sector of the
theory.

The fact, apparent from equations \gensol\ and \bans, that 
$b=a/2$ in this
solution makes its embedding in string
theory natural, as we will explain in the
next section.

\medskip

\noindent{\it $\Lambda\ne 0$, $a=0$}

In this case, the bulk equations of motion become
(in terms of $h\equiv \phi^\prime$ and 
$g\equiv A^\prime$)
\eqn\lameqs{\eqalign{
&h^\prime+4hg=0\cr
&6g^2-{2\over 3}h^2+{1\over 2}\Lambda=0\cr
&3g^\prime+{4\over 3}h^2=0\cr
}
}
We can solve the second equation for $g$ in terms of
$h$, and then integrate the first equation to
obtain $h(x_5)$.  For $g\ne 0$ the third equation
is then automatically satisfied.  
We will not need detailed properties of the solution, so we will
not include it here.  
The solutions are more complicated than those of \S2.3. 
We are currently exploring under what conditions one can solve
the matching equations to obtain a wall with singularities
cutting off the $x_5$ direction on both sides \toappear. 
If such walls exist, they will also exhibit the self-tuning phenomenon
of \S2.3, since the dilaton zero mode can absorb shifts in $V$ and
doesn't appear elsewhere in the action.

\medskip

\newsec{Toward a String Theory Realization}

\subsec{$\Lambda = 0$ Cases}

Taking $\Lambda = 0$ is natural in string theory, since the tree-level
vacuum energy in generic critical closed string compactifications
(supersymmetric or not) vanishes.
One would expect bulk quantum corrections to correct $\Lambda$
in a power series in $g_s = e^{\phi}$.  However, the analysis of
\S2.3\ may still be of interest if the bulk corrections to $\Lambda$
are small enough.  This can happen for instance if the
supersymmetry breaking is localized in a small neighborhood of
the wall and the $x_5$ interval is much larger, or 
more generally if the supersymmetry breaking scale in 
bulk is small enough. 

\medskip
\noindent {\it General $f(\phi)$}
\medskip

The examples we have found in \S2\ which ``self-tune'' the 4d cosmological
constant to zero have $\Lambda = 0$ with a broad range of
choices for $f(\phi)$.  We interpret this as meaning
that quantum corrections to the brane tension, which would change
the form of $f$, do not destabilize the flat brane solution. 
The generality of the dilaton coupling
$f(\phi)$ suggests that our results should apply to a wide
variety of string theory backgrounds involving domain
walls.  We now turn to a discussion of some of
the features of particular cases.

\medskip
\noindent {\it D-branes} 
\medskip

In string theory, one would naively
expect codimension one D-branes (perhaps wrapping a piece of
some compact manifold) to have $f(\phi)$ given by a power
series of the form 
\eqn\dbrane{f(\phi) = e^{{5\over 3}\phi} \sum_{n=0}^{\infty}
c_{n} e^{n\phi}} 
The $c_0$ term represents the tree-level D-brane tension
(which goes like ${1\over g_{s}}$ in string frame).  The higher
order terms in \dbrane\ represent quantum corrections from the
Yang-Mills theory on the brane, which has coupling $g_{YM}^{2} = e^{\phi}$.

If one looks for solutions of the equations which arise with the choice 
\dilcho\ for $f(\phi)$ with positive $V$ and
$b=5/3$ (the tree level D-brane theory), then there are no solutions
with finite 4d Planck scale.   The constraints of \S2.3\ cannot be
solved to give a single wall with singularities on both sides cutting
off the length in the $x_5$ direction.  However, including quantum
corrections to the D-brane theory to get a more generic $f$ as in 
\dbrane, there is a constraint on the magnitude of
${\partial f\over \partial \phi}(\phi(0))$ divided by $f(\phi(0))$
which can be obeyed.  Therefore, one concludes that for our
mechanism to be at work with D-brane domain walls, the dilaton
$\phi$ must be stabilized away from weak coupling -- the loop
corrections to \dbrane\ must be important.

\medskip
\noindent {\it The Case $f(\phi) = V e^{{2\over3}\phi}$ and NS Branes}
\medskip

Another simple way to get models which could come out of string theory
is to
set $b=2/3$ in \dilcho, so 
\eqn\nsbrane{f(\phi) = V e^{{2\over 3}\phi}}  
Then \basicac\ becomes precisely the 
Einstein frame action that one would get from a ``3-brane'' in string
theory with a string
frame source term proportional to $e^{-2\phi}$.
In this case, $\phi$ can also naturally be identified with the string
theory dilaton.
This choice of $b$ is possible in solutions of the sort summarized in
result (I) in \S2.1.

However, after identifying $\phi$ with the string theory
dilaton, if we really want to make this specific choice for
$f(\phi)$ we would also like 
to find branes where it is natural to expect that 
quantum corrections 
to the brane tension (e.g. from gauge and matter
fields living on the brane) would shift $V$, but not change
the overall $\phi$ dependence of the source term.  This can only
happen if
the string coupling $g_s = e^{\phi}$
is $\it not$ the field-theoretic coupling parameter for the
dynamical degrees of freedom on the brane.

Many examples where this happens are known in string theory.  
For example, the NS fivebranes of type IIB and heterotic
string theory have gauge fields on their worldvolume whose
Yang-Mills coupling does not depend on $g_s$ \refs{\smallinst,\ntwo,
\nati}.
This can roughly be understood from the fact that the
dilaton grows to infinity down the throat of the solution, and 
its value in the asymptotic flat region away from this
throat is irrelevant to the coupling of the modes on the
brane.  Upon compactification, this leads to gauge couplings
depending on sizes of cycles in the compactification manifold
(in units of $\alpha'$) \refs{\ntwo,\kss}.  
For instance,
in \kss\ 
gauge groups which arise ``non-perturbatively'' in singular heterotic
compactifications (at less supersymmetric 
generalizations of the small instanton singularity
\smallinst) are discussed.
There, the 4d gauge couplings on a
heterotic NS fivebrane wrapped on a two-cycle go like 
\eqn\branec{g_{YM}^{2} \sim {\alpha^\prime \over R^2}} 

\noindent
Here $R$ is the scale of this 2-cycle in the compactification manifold. 
In \kss, this was used to interpret string sigma model worldsheet
instanton effects, which go like $e^{-{R^{2}\over \alpha^\prime}}$,
in terms of nonperturbative effects in the brane gauge group,
which go like $e^{-{8\pi^2\over g_{YM}^2}}$.  So this is a
concrete example in which nontrivial dilaton-independent 
quantum corrections 
to the effective action on the brane arise.
One can imagine analogous
examples involving supersymmetry breaking.  
In such cases, perturbative 
shifts in the brane tension due to brane worldvolume gauge dynamics
would be a series in $\alpha^\prime \over R^2$ and not $g_s =
e^{\phi}$.

In particular,
one can generalize such examples to cases where the branes are
domain walls in 5d spacetime (instead of space-filling in 4d
spacetime as in the examples just discussed), but 
where again the brane gauge coupling is not the string coupling.
Quantum corrections to the brane tension 
in the brane gauge theory then naturally 
contribute shifts

\eqn\shiftac{ e^{{2\over 3}\phi} V \rightarrow
e^{{2\over 3}\phi} (V + \delta V)}
\noindent
to the (Einstein frame) $b=2/3$ 
source term in \basicac, without changing its
dilaton dependence.  

Most of our discussion here has focused on the case where $\phi$
is identified with the string theory dilaton.  However, 
in general it is possible that some other string theory modulus
could play the role of $\phi$ in our solutions, perhaps for more
general values of $b$.

\medskip
\noindent {\it Resemblance to Orientifolds}
\medskip

In our analysis of the equations, we find solutions describing
a 4d gravity theory with zero cosmological constant if
we consider singular solutions and cut off the fifth
dimension 
at these singularities.  
The simplest versions of compactifications involving
branes in string theory also include defects in
the compactification which absorb the charge of
the branes and cancel their contribution to
the cosmological constant in four dimensions,
at least at tree level.  Examples of these defects 
include 
orientifolds (in the context of D-brane worlds),
S-duals of orientifolds (in the context of
NS brane worlds), and Horava-Witten ``ends of the
world'' (in the context of the strongly coupled
heterotic string).  

Our most interesting solutions involve two different
behaviors on the two sides of the domain wall.  On
one side the dilaton goes to strong coupling while
on the other side it goes to weak coupling at the
singularity.  This effect has also been seen in
brane-orientifold systems \polwitt.  

It would be
very interesting to understand whether the singularities
we find can be identified with 
orientifold-like defects, as these similarities
might suggest.
Then their role (if any) in absorbing quantum corrections
to the 4d cosmological constant could be related to    
the effective negative tension of these defects.
However, various aspects of our dilaton gravity
solutions are not familiar from brane-orientifold
systems.  In particular, the existence of 
solutions with curved 4d geometry on the same footing as
our flat solutions does not occur in typical
perturbative string compactifications.  
In any case, note that (as explained in \S3.1) our mechanism does
not occur in the case of weakly coupled D-branes
and orientifolds.       

\subsec{$\Lambda \neq 0$ Cases}

Some of the $\Lambda \neq 0$ cases discussed in \S2.4\ could also
arise in string theory.  As discussed in \refs{\oldjoe,\kks} one can
find closed string backgrounds with nonzero tree level cosmological
constant $\Lambda < 0$ by considering subcritical strings.
In this case, the cosmological term would have dilaton dependence
consistent with $a = 4/3$ in bulk.  Using equations
\gensol\ and \bans, this implies $b = 2/3$, which
is the expected scaling for a tree-level brane tension in the
thin-wall approximation as well. 

One would naively expect to obtain vacua with such negative bulk cosmological
constants out of tachyon condensation in closed string theory
\refs{\oldjoe,\kks}.  It is  
then natural to consider these domain walls (in the $a=4/3, b=2/3$ case) 
as the thin wall approximation to ``fat'' domain walls which could
be formed by tachyon field configurations which interpolate between
different minima
of a closed string tachyon potential.  In the context of
the Randall-Sundrum scenario, such ``fat'' walls were studied
for example in \refs{\dfgk,\gremm,\csaki}.  

It would be interesting to find cases where the $\Lambda \neq 0$,
$a = 0$ solutions arise from a more microscopic theory.  
However, it is clear that the dilaton dependence
of \basicac\ is then not consistent with interpreting $\phi$ as the
string theory dilaton.  Perhaps one could find a situation where 
$\phi$ can be identified with some other string theoretic modulus, 
and $\Lambda$ can be interpreted as the bulk cosmological constant after
other moduli are fixed.

\newsec{Discussion}

The concrete results of \S2\ motivate many interesting questions,
which we have only begun to explore.  Answering these
questions will be important for understanding the
four-dimensional physics of our solutions.  

The most serious question has to do with the nature of the
singularities.  There are many singularities in string
theory which have sensible physical resolutions,
either due to the finite string tension or due to
quantum effects.  Most that have been studied (like
flops \flop\ and conifolds \conifold) involve systems
with some supersymmetry, though some (like orbifolds
\orbifold) can be understood even without 
supersymmetry.  We do not yet know the proper
interpretation of our singularities, though
as discussed in \S3\ there are intriguing similarities
to orientifold physics in our system.  After
finding the solutions, we cut off the volume integral
determining the four-dimensional Planck scale at
the singularities.  It is important to determine whether
this is a legitimate operation.

It is desirable (and probably necessary in order
to address the question in the preceding paragraph)
to embed our solutions microscopically into M theory.
As discussed in \S3, some of our solutions appear
very natural from the point of view of 
string theory, where the scalar
$\phi$ can be identified with the dilaton.  
It would be interesting to consider the analogous
couplings of string-theoretic moduli scalars
other than the dilaton.  Perhaps there are other
geometrical moduli which couple with 
different values of $a$ and $b$ in \dilcho\ than the
dilaton does.  

It is also important to understand the effects of
quantum corrections to quantities other than
$f(\phi)$ in our Lagrangian.  In particular, corrections
to $\Lambda$ and corrections involving different
powers of $e^\phi$ in the bulk (coming from loops of bulk
gravity modes) will change the nature of the equations.
It will be interesting to understand the details of 
curved 4d domain
wall solutions to the corrected equations \refs{\kaloper,\dfgk,\toappear}.  
More specifically, it will be of interest to
determine the curvature scale of the 4d slice,
in terms of the various
choices of phenomenologically natural
values for the Planck scale.
Since the observed value of the cosmological
constant is nonzero according to 
studies of the mass density, cosmic microwave
background spectral distribution, and supernova
events \perletc, such corrected
solutions might be of physical interest.     

Perhaps the most intriguing physical
question is what happens from the
point of view of four-dimensional effective
field theory (if such a description
in fact exists).  Understanding the singularity
in the 5d background is probably required to
answer this question.  One possibility (suggested
by the presence of the singularity and by
the self-tuning of the 4d cosmological constant
discovered here) is that four-dimensional
effective field theory breaks down in this background,
at least as far as contributions to the 4d cosmological
constant are concerned.  
In \RS\ and analogous examples, there is a continuum of
bulk modes which could plausibly lead to a breakdown of
4d effective field theory in certain computations.  In our
theories, cutting off the 5d theory at the singularities
leaves finite proper distance in the $x_5$
direction.  This makes it unclear how such a continuum 
could arise (in the absence of novel physics at the
singularities, which could include ``throats'' 
of the sort that commonly arise in brane solutions).    
So in this system, any 
breakdown of 4d effective field theory is more mysterious.

\medskip
\centerline{\bf{Acknowledgements}}
\medskip

We are indebted to O. Aharony and T. Banks for interesting
discussions which motivated us to investigate this subject.  We would
also like to thank 
R. Sundrum for many helpful discussions about
closely related topics; we understand that Arkani-Hamed, 
Dimopoulos, Kaloper
and Sundrum have uncovered very similar results \nnr.  
We thank H. Verlinde for interesting discussions,
and in particular for several helpful comments about
the potential generality of these results.
In addition
we are grateful to M. Dine, N. Kaloper, S. Shenker, M. Shmakova, 
L. Susskind and 
E. Verlinde
for stimulating discussions. 
We would like to acknowledge the kind hospitality of the School of
Natural Sciences at the Institute for Advanced Study during the 
early stages of this work.  S.K. is supported in part by
the Ambrose Monell Foundation and a 
Sloan Foundation Fellowship, M.S. is supported in part by 
an NSF Graduate Research Fellowship, and E.S. is supported in part by
a DOE OJI Award and a Sloan Foundation Fellowship.  S.K. and
E.S. are supported in part by the DOE under contract
DE-AC03-76SF00515.

\listrefs
\end